\documentstyle[aps,prl,preprint]{revtex}
\input psfig
\draft

\tighten

\begin{document}

\title{Effects of a nano-sized filler on the structure and dynamics 
of a simulated polymer melt and the relationship to ultra-thin films}

\author{Francis~W. Starr, Thomas B. Schr{\o}der, and Sharon C. Glotzer}

\address{Polymers Division and Center for Theoretical and Computational
Materials Science, National Institute of Standards and Technology,
Gaithersburg, Maryland 20899, USA}

\date{19 July 2000}
 
\maketitle

\begin{abstract}
We perform molecular dynamics simulations of an idealized
polymer melt 
surrounding a nanoscopic filler particle to probe the effects of a
filler on the local melt structure and dynamics.  We show that the glass
transition temperature $T_g$ of the melt can be shifted to either higher
or lower temperatures by appropriately tuning the interactions between
polymer and filler.  A gradual change of the polymer dynamics
approaching the filler surface causes the change in the glass
transition.  We also find that while the bulk structure of the polymers
changes little, the polymers close to the surface tend to be elongated
and flattened, independent of the type of interaction we study.
Consequently, the dynamics appear strongly influenced by the
interactions, while the melt structure is only altered by the geometric
constraints imposed by the presence of the filler.  Our findings show a
strong similarity to those obtained for ultra-thin polymer films
(thickness $\lesssim 100$~nm) suggesting that both ultra-thin films and
filled-polymer systems might be understood in the same context.
\end{abstract}
\bigskip

Significant enhancements in mechanical, rheological, dielectric,
optical, and other properties of polymer materials can be obtained by
adding fillers such as carbon black, talc, silica, and other inexpensive
materials~\cite{wypych}.  Applications of filled polymers are diverse,
ranging from automobile tires and bumpers to the rapidly expanding area
of micro- and nano-electronic devices, which are continually challenged
by the progressive miniaturization of devices~\cite{wypych,nanotech}.
In particular, the growing ability to design customized nano-fillers of
arbitrary shape and functionality provides an enormous variety of
property modifications by introducing specific heterogeneity at the
nanoscale~\cite{nanotech,poss1,poss2}.  However, detailed knowledge of
the effects of fillers on a polymer melt at the molecular level is
lacking due to the difficulty of directly probing the polymer structure
and dynamics in the vicinity of the polymer-filler interface.  In this
regard, molecular simulations provide an ideal opportunity for direct
insight into filled materials.  Additionally, the understanding of
ultra-thin polymer films, which also have many important technological
applications (e.g. paints, lubricants, adhesives, and electronic
packaging), is a topic of continuing
discussion~\cite{karim-book,kumar-nature,kraus,theodorou,kumar-sim1,%
kumar-sim2,binder91,vanzanten,wallace,dutcher,forrest,lin}; the present
results provide a framework in which to interpret experiments on filled
polymers, and also possibly polymer thin films, which report both
increases and decreases of the glass transition temperature
$T_g$~\cite{angell95}, depending on the details of the system
studied~\cite{wypych,filled1,filled2,filled4}.
 
Our findings are based on extensive molecular dynamics simulations of a
single nanoscopic filler particle surrounded by a dense polymer melt
(Fig.~\ref{fig:pretty-pic}).  We simulate 400 chains of 20 monomers each
(below the entanglement length), a total of 8000 monomers.  The polymers
are modeled as chains of monomers, which interact via a Lennard Jones
(LJ) potential $V_{\mbox {\scriptsize LJ}}(r) =
4\epsilon((\sigma/r)^{12} - (\sigma/r)^6)$.  Additionally, bonded
monomers are connected via a FENE anharmonic spring potential $V_{\mbox
{\scriptsize FENE}} = -k(R_0^2/2) \ln
(1-(r/R_0)^2)$~\protect\cite{fene1,fene3}.  Several authors have studied
the pure (unfilled) melt ~\protect\cite{benneman1,benneman2}, and the
system has been shown to be a good glass-former.  This type of
``coarse-grained'' model is frequently used to study general trends of
polymer systems, but does not provide information for a specific
polymer.  The filler particle is modeled as an icosahedron, not unlike
primary particles of graphitized carbon
black~\protect\cite{carbon-black}.  We assign ideal force sites at the
vertices, at 4 equidistant sites along each edge, and at 6 symmetric
sites on the interior of each face of the icosahedron, as shown in the
right panel of Fig.~\ref{fig:pretty-pic}.  We tether a particle to each
of these sites by a FENE spring, which maintains a relatively rigid
structure but allows for thermalization of the
filler.~\cite{filler-note}.  We consider a filler particle with an
excluded volume interaction only, as well as one with excluded volume
plus attractive interactions (expected for many fillers), to determine
which properties are a result of the steric constraints imposed by the
filler, and which properties are affected by polymer-filler attraction.
We choose the same parameters for the interaction potential for all
filler force sites.  Periodic boundary conditions are used in all
directions.

Our model filler has several general features typical of a primary
carbon black particle (a traditional filler)~\cite{wypych,carbon-black},
as well as some newer nano-fillers~\cite{poss1,poss2}: (i) it has a size
of order 10~nm and (ii) it is highly faceted, but nearly spherical.  The
size of the facets is roughly equal to the end-to-end distance $R_e$ of
the low molecular weight polymers comprising the surrounding melt (for
Gaussian chains, $R_e^2 = 6 R_g^2$, the radius of gyration).  Since we
aim to study the changes induced by a single filler on the nanoscopic
level, and the implications of those changes for bulk properties, we
also consider a pure dense melt for comparison.

\begin{figure}[htbp]
\begin{center}
\mbox{\psfig{figure=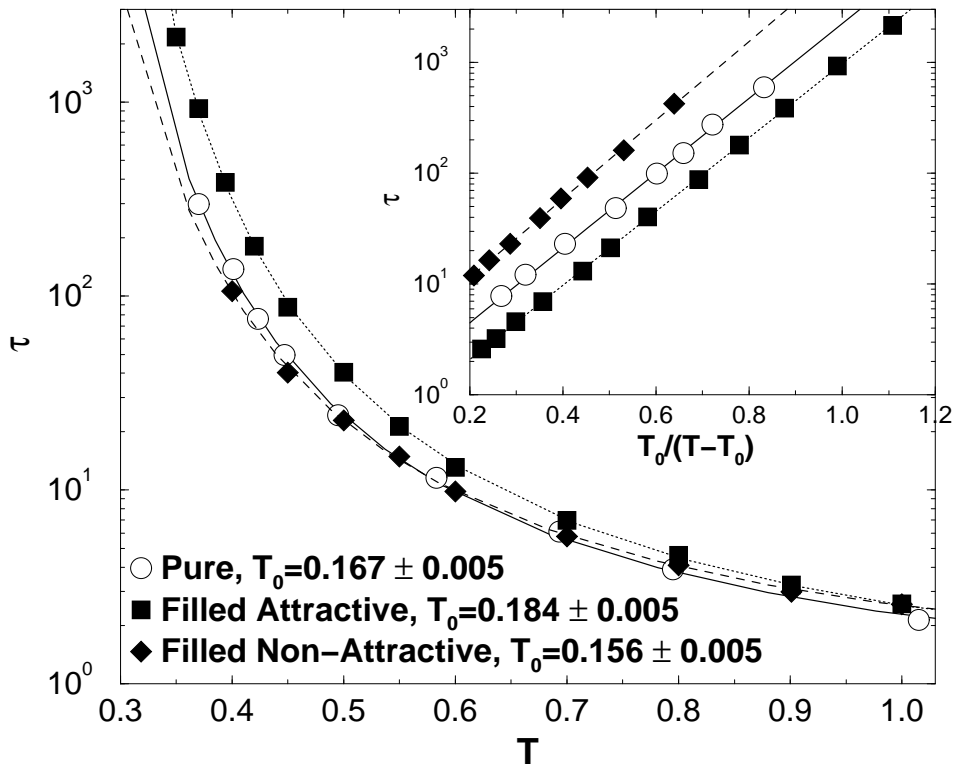,angle=-90,width=5in}}
\end{center}
\caption{Temperature dependence of the relaxation time of the
collective intermediate scattering function.  For attractive
polymer-filler interactions, $\tau$ is increased at all $T$ relative to
the pure system, and correspondingly has a greater value of $T_0$ (and
thus $T_g$).  Conversely, $\tau$ is reduced and $T_0$ is smaller for
non-attractive (excluded volume only) interactions.  The lines are a fit
to the VFT form.  The inset shows the same data plotted against reduced
temperature $T_0/(T-T_0)$ to show the quality of the VFT fit.  For
clarity of the inset, $\tau$ of the pure system is multiplied by 2, and
$\tau$ of the filled non-attractive system is multiplied by 4.
}
\label{fig:tau-isf}
\end{figure}

We simulate the pure system at density $\rho = 1.0$ at temperatures
ranging from $T=0.37$ -- $1.0$~\cite{units}.  We simulate the filled
systems in the range $T=0.35$ -- $1.2$.  Equilibration times range from
$5\times10^2 \tau^*$ at the highest $T$ to $2 \times 10^4 \tau^*$
(approximately 40~ns in Argon units) at the lowest $T$; we use the
rRESPA multiple time step algorithm to improve simulation
speed~\cite{respa,therm-note}.  In order to compare the simulations of
the filled system with the pure melt, we choose the box size so that the
local density far from the filler deviates at most by 0.2\% from the
density of the pure melt; such a density difference would cause a change
in $T_g$ in this model less than that reported in
Fig.~\ref{fig:tau-isf}~\cite{benneman1}.  For attractive monomer-filler
interactions, a box size $L=20.4$ (all lengths in units of
$\sigma_{\mbox {\scriptsize mm}}$) satisfies this constraint at all $T$.
In the non-attractive case, the characteristic first neighbor distance
between the filler sites and monomers is $T$ dependent due to the lack
of a unique minimum in the polymer-filler interactions.  As a result, at
each $T$ a different $L$ is required to achieve the correct $\rho$ at
large distance from the filler.  The box sizes range from $L = 20.49$ at
$T=1.0$, to $L=20.6$ at $T=0.4$.

To quantify the effect of the filler on $T_g$ --- one of the most
important processing parameters --- and on dynamic properties, we first
calculate the relaxation time $\tau$ of the intermediate scattering
function

\begin{equation}
F(q,t) \equiv \frac{1}{NS(q)} \left\langle \sum_{j,k=1}^N e^{-i {\bf
q}\cdot[{\bf r}_k(t) - {\bf r}_j(0)]} \right\rangle ,
\label{eq:isf}
\end{equation}

\noindent where the sum is performed over all $N$ monomers and we
normalize by the structure factor $S(q)$ such that $F(q,0) = 1$.  We
define the value of $\tau$ by $F(q,\tau) \equiv 0.2$.  Relative to the
pure system, we find that $\tau$ is larger at each $T$ for the
attractive system (Fig.~\ref{fig:tau-isf}).  Conversely, $\tau$ appears
slightly smaller at low $T$ in the non-attractive system, but is nearly
indistinguishable from the results for the pure system.
The lines are fits to the Vogel-Fulcher-Tammann (VFT) form~\cite{pablo}

\begin{equation}
\tau \sim e^{A/(T-T_0)}
\label{eq:vft}
\end{equation}

\noindent where $T_0$ is typically quite close to the experimental $T_g$
value~\cite{pablo}; hence changes in $T_g$ are reflected in $T_0$.
Consistent with the changes in $\tau$ relative to the pure melt, we find
that $T_0$ increases in the system with attractive interactions, but
clearly decreases in the system with only an excluded volume
interaction.  Thus, the effect of the steric hindrance introduced by the
filler particle decreases $\tau(T)$ and $T_g$, in spite of the fact that
monomers have a reduced number of directions in which to move, and hence
degrees of freedom that aid in the loss of correlations.  The fact that
$T_g$ shifts in opposite directions for attractive versus purely
excluded volume interactions demonstrates the importance of the surface
interactions.

\begin{figure}[htbp]
\begin{center}
\mbox{\psfig{figure=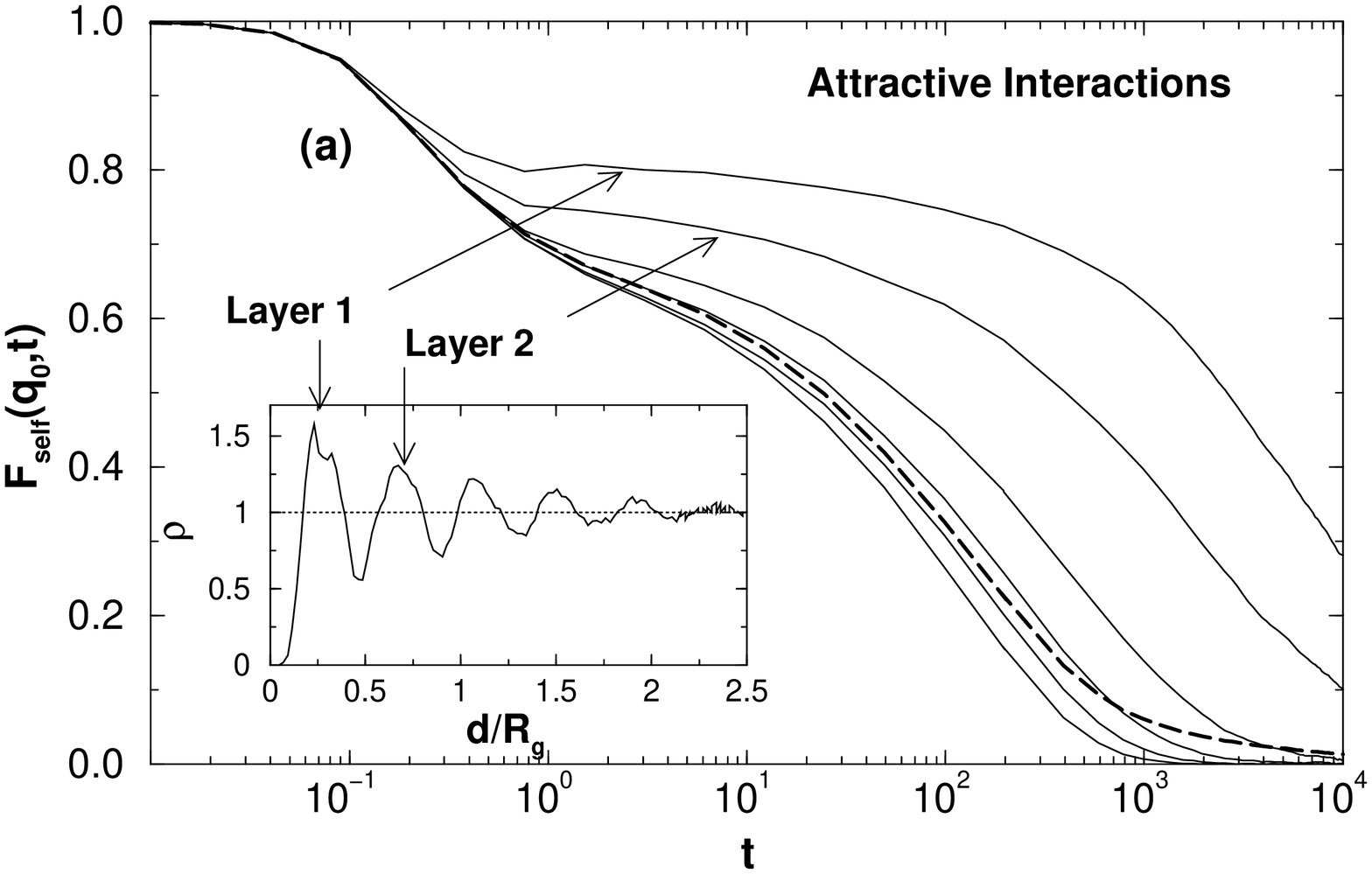,angle=-90,width=5in}}
\\
\mbox{\psfig{figure=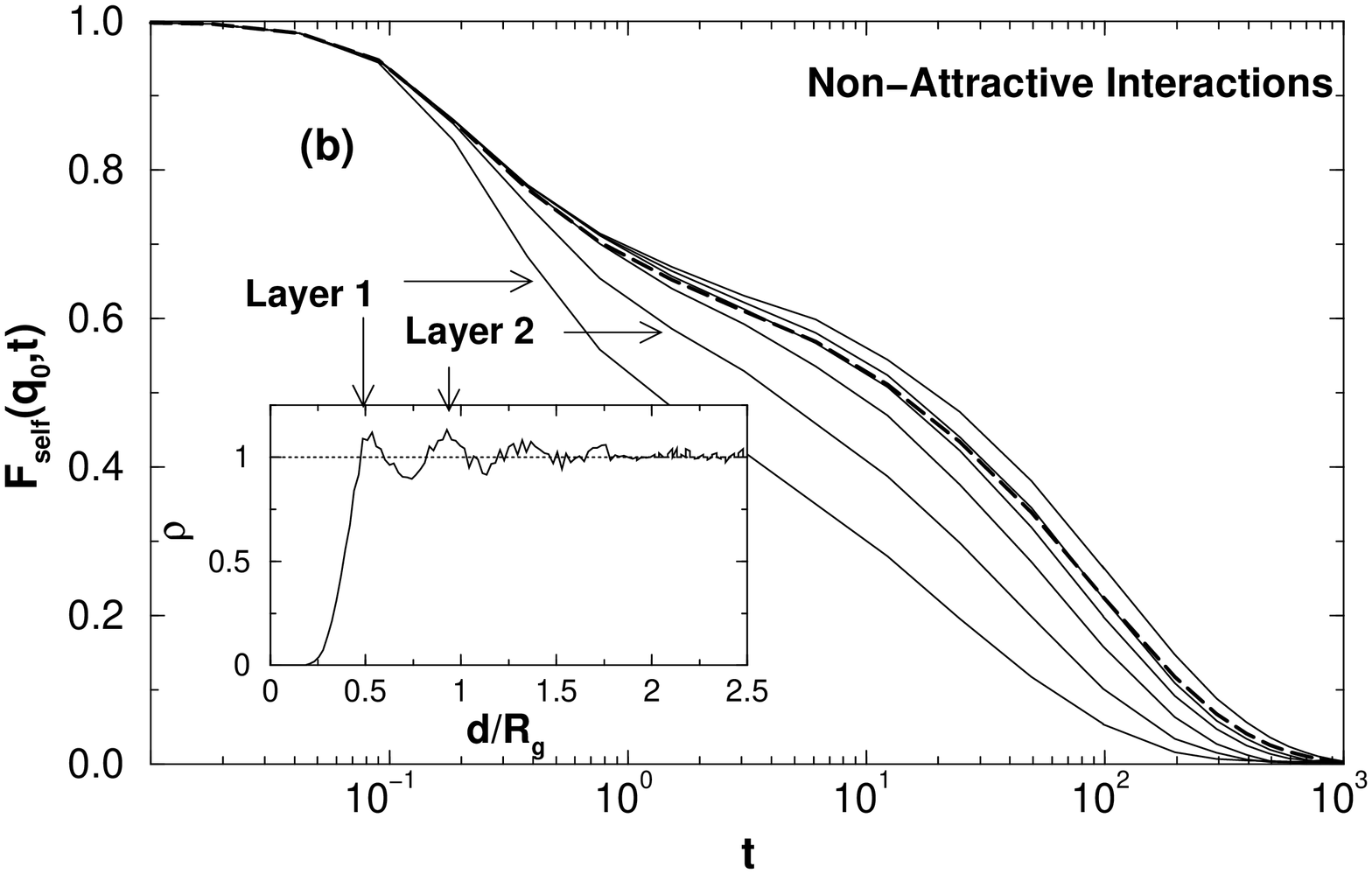,angle=-90,width=5in}}
\end{center}
\caption{The self (incoherent) part of the intermediate scattering
function for the average of all monomers (dotted line) and decomposed
into layers (defined by the distance from the filler surface) for {\bf
(a)} attractive interactions and {\bf (b)} non-attractive interactions
at temperature $T=0.4$.  The inset of each figure shows the local
density profile $\rho(d/R_g)$ of monomers as a function of distance from
the filler, normalized by $R_g$ of the polymers.  We define the distance
$d$ from the filler surface as the difference between the radial
position of a monomer $r_{\mbox{\scriptsize mon}}$ and the radius of the
inscribed sphere of the icosahedral filler particle
$r_{\mbox{\scriptsize icos}} = \frac{1}{12} (42+18
\protect\sqrt{5})^{\frac{1}{2}}L$, where $L$ is the length of an edge of
the icosahedron.  The monomers order in well-defined layers surrounding
the filler; we use the minima in $\rho(r)$ to define the boundary
between layers.  The main part of each figure shows
$F_{\mbox{\scriptsize self}}(q_0, t)$ for each of the layers.  We
calculate $F_{\mbox{\scriptsize self}}(q_0,t)$ --- given by
Eq.~\ref{eq:isf} with the sum restricted to only one index --- for
monomers located in each layer at $t=0$.  In {\bf (a)}, we see that the
relaxation near the filler surface is slowed by roughly 2 orders of
magnitude.  This is consistent with the dynamics being slower in the
filled attractive system (and hence an increased $T_g$).  In contrast,
in {\bf (b)} the relaxation of $F_{\mbox{\scriptsize self}}(q_0, t)$ is
enhanced near the surface.  This is consistent with the dynamics being
faster in the filled non-attractive system (and hence a decreased
$T_g$).  The relaxation time of the outer most layer in both cases
nearly coincides with the relaxation time of the pure system.  }
\label{fig:fself-layers}
\end{figure}

To elucidate how the local dynamics of the monomers are influenced by
the filler, we examine the relaxation of the self (incoherent) part
$F_{\mbox{\scriptsize self}}(q,t)$ of $F(q,t)$ as a function of the
monomer distance from the filler.  Monomers typically form layers near a
surface~\cite{binder91}; we find well-defined monomer layers surrounding
the filler, as seen in the density profile of
Fig.~\ref{fig:fself-layers}.  Hence, we split $F_{\mbox{\scriptsize
self}}(q,t)$ into the contribution from each separate layer.
Specifically, we calculate $F_{\mbox{\scriptsize
self}}^{\mbox{\scriptsize layer}}(q,t)$ using the monomers located in
each layer at $t=0$ such that $F_{\mbox{\scriptsize self}}(q,t) = 1/N
\sum_{\mbox {\scriptsize layers}} N_{\mbox {\scriptsize layer}}
F_{\mbox{\scriptsize self}}^{\mbox{\scriptsize layer}}(q,t)$, where
$N_{\mbox{\scriptsize layer}}$ is the number of monomers in a given
layer.  We show $F_{\mbox{\scriptsize self}}^{\mbox{\scriptsize
layer}}(q_0,t)$, as well as $F_{\mbox{\scriptsize self}}(q_0,t)$ for one
temperature in Fig.~\ref{fig:fself-layers}.  In the attractive system,
the relaxation of the layers closest to the filler are slowest,
consistent with the system dynamics being slowed by the attraction to
the filler.  Conversely, for the non-attractive system, we find that the
relaxation of inner layer monomers is significantly enhanced compared to
the bulk, consistent with the observed enhancement of the system
dynamics.  Preliminary results support the possibility that faster
dynamics may also occur with attractive interactions, provided that the
polymer-filler attraction is weaker than that of polymer-polymer
interactions.  The altered dynamics persist for a distance slightly less
than $2 R_g$ from the surface.  Our results demonstrate that
interactions play a key role in controlling $T_g$ and the local dynamics
of filled polymers.  We expect the role of interactions to be largely
the same when many filler particles are present in the melt, but there
will be additional effects on dynamic properties due to the more complex
geometrical constraints.

\begin{figure}[htbp] 
\begin{center}
\mbox{\psfig{figure=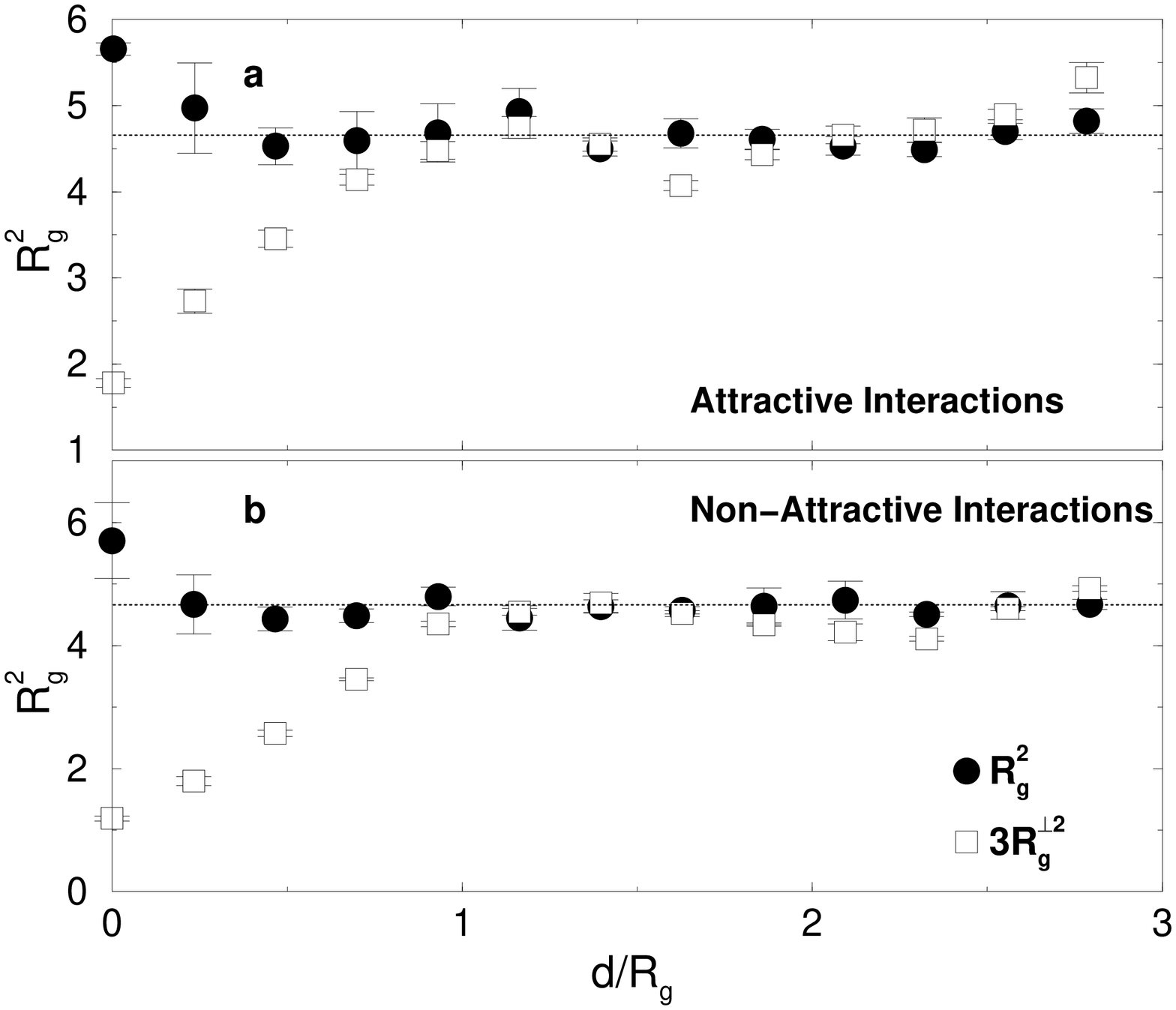,angle=-90,width=5in}}
\end{center}
\caption{Radius of gyration $R_g$ of the polymer chains as a function of
distance $d/R_g$ of the center of mass of a chain from the filler
surface for $T=0.4$.  We also resolve the component perpendicular to the
surface, which we label by $R_g^\perp$.  We show results for {\bf (a)}
attractive and {\bf (b)} non-attractive interactions.  The increase of
$R_g$ as the surface of the filler is approached indicates that the
chains become increasingly elongated.  The decrease of $R_g^\perp$ shows
that the chains also become ``flattened'' in the radial direction --
roughly perpendicular to the filler surface.  The effect appears largely
independent of the temperature and numerical values of the potential
parameters.  }
\label{fig:radii} 
\end{figure}

We next turn our attention to any structural effect the filler has on
the melt.  Quantities such as the pair distribution function,
average $R_g$, average $R_e$, and distribution of bond lengths and
angles show no significant deviations from the pure system.  However,
by focusing on the dependence of $R_g$ (or $R_e$) on the distance $d$
from the filler surface, we find a change in the overall polymer
structure near the surface.  In Fig.~\ref{fig:radii}, we show $R_g^2$,
as well as the radial component from the filler center $R_g^{{\small
\perp} 2}$ (approximately the component perpendicular to the filler
surface) for both attractive and non-attractive polymer-filler
interactions at one temperature.  $R_g^2$ increases by about 50\% on
approaching the filler surface; at the same time $R_g^{{\small \perp}
2}$ decreases by slightly more than a factor of 2 for both attractive
and non-attractive systems.  The combination of these results indicates
that the polymers become slightly elongated near the surface, and
flatten significantly.  Note that not all monomers belonging to a given
``surface polymer'' are located in the first surface layer, as depicted
in Fig.~\ref{fig:pretty-pic}.  We also point out that the chains retain
a Gaussian conformation near the filler surface~\cite{gauss-note}.  We
find that the range of the flattening effect roughly spans a distance
$R_g$ from the surface, and the results depend only weakly on $T$.  We
performed an additional simulation with double the attraction strength
between the filler and polymers and did not find any significant further
effect on the chain structure.  The independence of the chain structure
on the interaction suggests that the altered shape of the polymers is
primarily due to geometric constraints of packing the chains close
($d\lesssim R_g$) to the surface.  For significantly stronger
interactions, alteration of the chain structure is expected on
theoretical grounds~\cite{jack}.

We next consider the implications that our results may have for studies
of ultra-thin polymer films, where there is long standing debate on the
role of interactions versus confinement on $T_g$
shifts~\cite{karim-book,vanzanten,wallace}, local melt
dynamics~\cite{karim-book,dutcher,forrest,lin}, and melt
structure~\cite{karim-book,kumar-nature,kraus,theodorou,kumar-sim1,%
kumar-sim2,binder91}.  Our simulations allow us to address the effects
of interactions with a surface, without the additional complication of
confinement effects present in thin films.  It is largely agreed that
ultra-thin films with strongly attractive substrates increase $T_g$,
while weak substrate interactions (or no substrate, as in freely
standing films) lead to a downward shift of $T_g$; this is consistent
with our results.  This consistency is reasonable for fillers which have
facets that are relatively smooth and large compared to $R_g$; for
nanoscopic fillers, such as we study, it is surprising that a
correspondence occurs even for $R_g$ close to the filler size.  Not
surprisingly, the magnitude of the shifts we observe depends on the
relative quantities of polymer and filler; a greater filler
concentration would have a more dramatic effect (as observed
experimentally in refs.~\cite{filled1,filled2,filled4}).  Insofar as the
magnitude of effects depends only on the ratio of the surface to bulk
monomers, the thickness of the film is analogous to the inverse of the
concentration of the filler.  This is consistent with the experimental
observation that $T_g$ shifts are more pronounced as film thickness
decreases.  Recently there have been several experiments on segmental
motion in both freely-standing and supported ultra-thin
films~\cite{karim-book,dutcher,forrest}.  The observed segmental
dynamics is consistent with a decreased $T_g$ find in calorimetric
measurements~\cite{karim-book,vanzanten,wallace}.  At this time, it is
not clear whether a model with layers of different mobility is
applicable to understand $T_g$ shifts of thin films~\cite{karim-book};
however, the parallel behavior we observe between the thin films and our
simulations of a filled melt support this viewpoint.  Finally, the
elongation and flattening of polymers we observe near the filler has
been observed in thin-film
simulations~\cite{theodorou,kumar-sim1,kumar-sim2,binder91} as well as
recent experiments ~\cite{kumar-nature,kraus}; the range of the effect
found in Ref.~\cite{kumar-nature} is quantitatively consistent with our
results, which show the effect only for a range of roughly $R_g$, while
the results of Ref.~\cite{kraus} observed flattening for film
thicknesses $\lesssim 6 R_g$.  We also found, as in
Ref.~\cite{kumar-nature}, that the chains retain a Gaussian structure
near the surface.  Thus our findings demonstrate that confinement is not
a necessary ingredient for the observed changes in the dynamics and
structure of polymers near surfaces.  While our results provide strong
support for interpreting the results for filled melts and ultra-thin
films in the same framework, it is obvious that much care must be used
when analyzing specific systems.

We wish to thank E.~Amis, J.~Douglas, Y.~Gebremichael, C.~Han, A.~Karim,
N.~La\v{c}evi\'c, and A.~Nakatani, and W.~Wu.  We also thank the High
Performance System Usage group at NIST for use of the SGI Origin
Supercomputer.  F.W.S. thanks the National Research Council for support.

\begin{figure}[htbp]
\begin{center}
\end{center}
\caption{[Figure excluded because of file size].  ``Snapshot'' of our
simulation of the filled polymer melt.  The bonds between
nearest-neighbor monomers along a chain are drawn in various shades of
gray for clarity.  The left panel shows the filler with the surrounding
melt; the right panel shows a few representative polymers that have
monomers near the filler surface.  }
\label{fig:pretty-pic}
\end{figure}

\begin{references}
\bibitem{filler-note} Relative to a monomer, the mass of a filler site
$m_{\mbox{\scriptsize f}} = 2m_{\mbox{\scriptsize m}}$, the spring
stiffness $k_{\mbox{\scriptsize f}} = 1.5k_{\mbox{\scriptsize m}}$, and
the bond parameter $R_0^{\mbox{\scriptsize f}} =
2/3R_0^{\mbox{\scriptsize m}}$.  We choose a stiffer spring, larger
mass, and smaller bond parameter than used for the polymers to reduce
the fluctuation of filler sites.  In order to avoid any spurious effects
that might be induced by a large cavity inside the filler particle, a
replica of the particle with three sites on an edge and one with only
sites at the vertices are layered inside the outer icosahedron,
reminiscent of a ``Russian doll''.  This yields a total 356 force sites
associated with the filler.  For the attractive case, we use an LJ
particle with an interaction strength between filler sites
$\epsilon_{\mbox{\scriptsize ff}} = 2.0 \epsilon_{\mbox{\scriptsize
mm}}$.  We use the Lorentz-Berthelot~\cite{allen-tildesley} mixing rules
for the monomer-filler interactions and take
$\epsilon_{\mbox{\scriptsize mf}} = \sqrt{\epsilon_{\mbox{\scriptsize
mm}} \epsilon_{\mbox{\scriptsize ff}}}$.  For the non-attractive case,
we drop the attractive $r^6$ term in the LJ potential for monomer-filler
interactions and retain only the $r^{12}$ term, accounting for excluded
volume.  For some fillers, it is expected that interactions with the
filler edges will be stronger than interactions with the face.  It is
not immediately clear how the qualitative results for such a filler
might differ
\bibitem{units} All values are reported in reduced LJ units.
Standard units for $T$ are recovered by multiplying $T$ by
$\epsilon_{\mbox {\scriptsize mm}}/k_B$, where $k_B$ is Boltzmann's
constant.  Time is given by $\tau^* = (m_{\mbox {\scriptsize
m}}\sigma_{\mbox {\scriptsize mm}}^2 / \epsilon_{\mbox {\scriptsize
mm}})^{1/2}$.
\bibitem{therm-note}
The temperature is adjusted using the Andersen
method~\cite{allen-tildesley}; for production runs, the temperature is
weakly coupled to the Berendsen thermostat~\cite{allen-tildesley} to
prevent a drift in the total energy due to the accumulation of round-off
error.
\bibitem{fqt-note} The exact choice for the definition of $\tau$ does
not significantly change the temperature dependence of our results.  We
focus our attention on the magnitude of wave-vector $q_0\approx7.08$,
the location of the first maximum in $S(q)$.
\bibitem{gauss-note} The chains retain a Gaussian confirmation in the
sense that $R_e^2 = 6R_g^2$; however, the probability distribution of
the monomer coordinates is not symmetric in all directions.
\end{references}
\end{document}